# *Implementation of the Six Channel Redundancy to achieve fault tolerance in testing of satellites.*

H S Aravinda
Dept of ECE, REVA ITM,
Bangalore-64, Karnataka, India
aravindhs1@gmail.com

Dr H D Maheshappa
Director & Principal, E PC ET
Bangalore- 40, Karnataka, India
hdmappa@gmail.com

Dr Ranjan Moodithaya
Head, KTM Division, NAL,
Bangalore-17, Karnataka, India
ranjanmk@hotmail.com

*Abstract:-***This paper aims to implement the six channel redundancy to achieve fault tolerance in testing of satellites with acoustic spectrum. We mainly focus here on achieving fault tolerance. An immediate application is the microphone data acquisition and to do analysis at the Acoustic Test Facility (ATF) centre, National Aerospace Laboratories. It has an 1100 cubic meter reverberation chamber in which a maximum sound pressure level of 157 dB is generated. The six channel Redundancy software with fault tolerant operation is devised and developed. The data are applied to program written in C language. The program is run using the Code Composer Studio by accepting the inputs. This is tested with the TMS 320C 6727 DSP, Pro Audio Development Kit (PADK).**

Key words: Fault Tolerance, Redundancy, Acoustics.

## I. INTRODUCTION

Acoustic Test Facility is a national facility for acoustic environmental qualification of satellites, launch vehicle stages and their subsystems for the ISRO [1]. The ATF has a reverberation chamber (RC) for simulating the acoustic environment experienced by spacecraft and launch vehicles during launch [2]. The RC has a diffused uniform sound pressure level distribution. Its wall surface ensures reflectance of 99% of the sound energy. It is used for simulating the acoustic environment experienced by spacecraft and launch vehicles during the launch. The one such facility is shown in Fig.1.

The Indian Space Research Organization launches number of satellites for application in communication [5], remote sensing, meteorology etc. The powerful launch vehicles are used to accelerate the satellite through the earth's atmosphere and to make it an artificial earth satellite. The Launch Vehicles [6] used will generate high levels of sound during lift-off and Tran's atmospheric acceleration. The payload satellites experiences mechanical loads of various frequencies and load on the vehicle from acoustic sources due to two factors. One is Rocket vehicle generated noise at lift-off, and the other is an aerodynamic noise caused by turbulence, particularly at frontal area transition. The acoustic field thus created is strong enough to damage the delicate payload. The sources of acoustics, its combined spectrum are shown in fig.2 and fig.3.

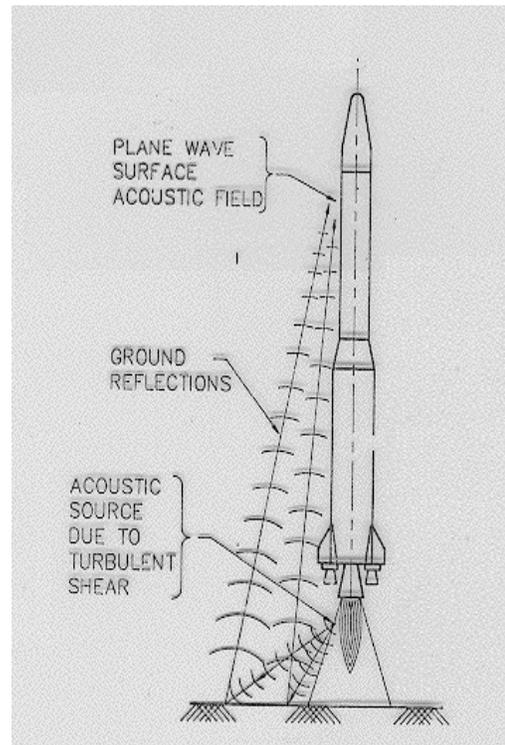

Fig.2. the load on the vehicle from two acoustic sources.

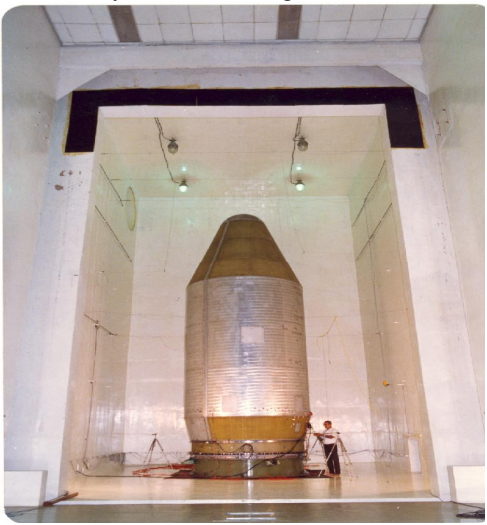

Fig.1. View of the Reverberation Chamber

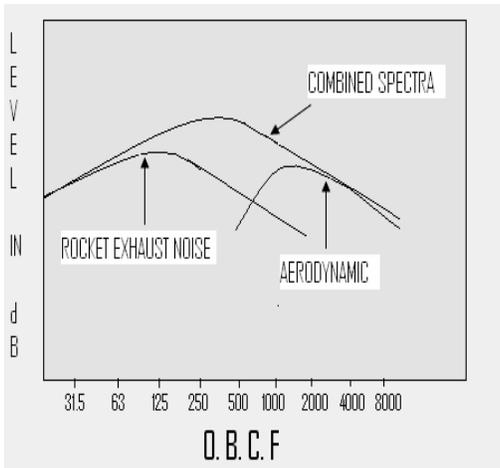

Fig.3. The combined spectra from the two acoustic sources.

Faults and failures are not acceptable due to high cost of satellites. Hence, all payload satellites should undergo an acoustic test before launching under simulated conditions and are tested for their endurance of such dynamic loads. The satellite is subjected to maximum overall sound pressure level to ensure the functional aspects of all the test setup. Acoustic test is a major dynamic test for qualification of space systems and components. The purpose of the tests are, Search for weak elements in the subsystem with respect to acoustic failure. The Qualification tests to demonstrate spacecraft performance in meeting design goals set. Acceptance tests to uncover workmanship nature of defects.

## II. ACOUSTIC TESTING

The acoustic environment inside the Reverberation Chamber is created by modulating a stream of clean and dry air at about 30 PSI pressure using electro pneumatic transducers. The drive signal is derived from a random noise generator and modified by a spectrum shaper. The microphone data from the RC is observed on a real time analyzer and the spectrum shaper outputs are adjusted to achieve the target spectrum. There are two sets of modulators, one delivering an acoustic power of 60KW in the 31.5 Hz to 500 Hz and the others delivering 20 KW in the 200 to 1200 Hz range, the spectrum beyond 1200 HZ is controlled to some extent using the effects of the higher harmonics by changing the spectral contents of the drive to the modulators. The acoustic excitation is coupled to the RC through optimally configured exponential horns to achieve efficient transfer of the acoustic energy into the chamber. The chamber wall surface treatment design ensures reflectance of 99% of the sound energy incident on them. The chamber has a diffused uniform, sound pressure level distribution with in ± 1dB in the central ten percent of volume of the chamber where the test specimen is located. The spectrum for almost all contemporary launch vehicles around the world can be realized in the RC, like the US Delta, Atlas Centaur, Titan IIIC, Space Shuttle, Ariane 4 & 5 of the ESA, Vostok, Soyuz of Russia and ASLV, PSLV and GSLV of India.

### A. Requirements For acoustic Testing of Satellite are

Noise generation unit, Spectrum Shapers, Power amplifiers and horns (two), Reverberation Chamber, Micro phones and Multiplexer, Real time frequency (spectrum) analyzer, Graphic recorder and display, Tape recorder for recording, Accelerometers, Charge amplifier and data recorder, Dual channel analyzer and plotter.

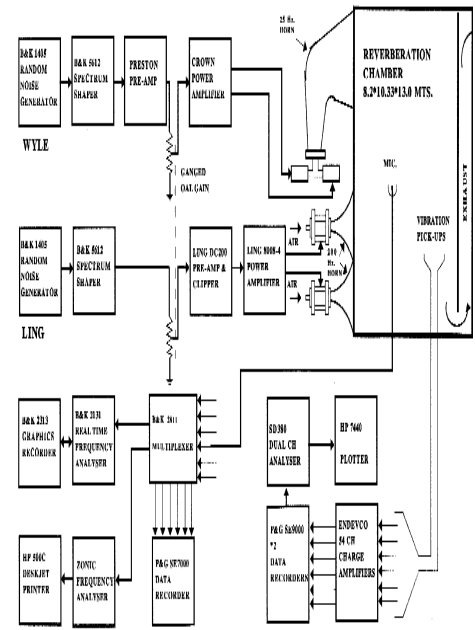

FIG.4 NOISE GENERATION, MEASUREMENT AND DATA ACQUISITION SYSTEM

The satellite is kept in RC and the high frequency high level spectrum characteristics of the launch vehicle are generated and its dynamic behavior is studied. It is essential that the acoustic noise generated is a true simulation of the launch vehicle acoustic spectrum, and it is the input acoustic load to be simulated in the RC and is specified as SPL (sound pressure level) in dB verses frequency. The spectrum of various launch vehicles like delta, atlas centaur, titan-IIIC, Arianne, vostok, ASLV, PSLV, GSLV.., and Indian satellites like IRS, INSAT.., are realizable in the Reverberation Chamber. Each launch vehicle has unique spectral features and is drawn in Octave Band Centre Frequencies (OBCF), in the range from 31.5 Hz to 16 kHz.

The Three levels of acceptance of acoustic spectrum are , first is Full level or Qualification test, it is normally for 120 seconds and Maximum of 156dB, second is Acceptance level test, it is normally for 60 or 90 seconds and Maximum of 153dB, third is a Low level test, it is normally for 30 seconds and Maximum of 150dB.

### III. IMPLEMENTATION

Fault tolerant application software to ensure data integrity will be developed. This paper is implemented by taking the six channel data from reverberation chamber and is applied as the input to the program. The six microphones data are connected to the *TMS 320C 6727 DSP, Pro Audio Development Kit (PADK)* after signal conditioning via analog to digital converters. All six microphone data is fed to DSP processor as shown in Fig.5. The FFT is taken for all the six channel data and are compared with each other to find out which channel microphone data is good or bad. A threshold level is maintained to check the validity of the microphone. If the data is well with in the threshold it is accepted or else it will be rejected. Here if the two channel microphone data is bad then it will only be identified.

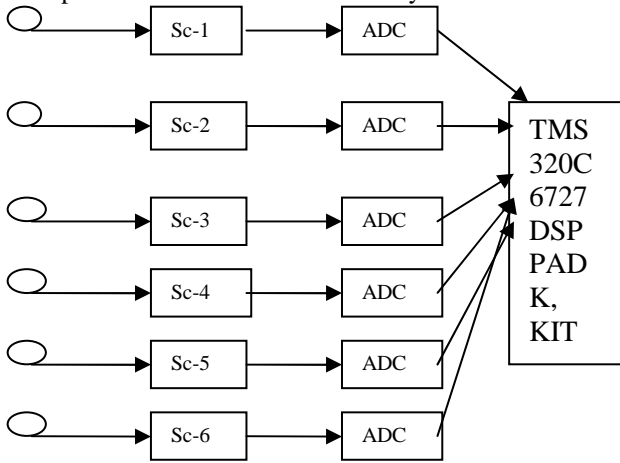

Fig.5. Block Diagram representation for six channel redundancy management technique.

### IV. TEST AND RESULTS

The data is extracted using the six microphone channels. It is fed to the TMS320C 6727 DSP, *Pro Audio Development Kit (PADK) for further processing* using the Code Composer Studio for different cases. The data is applied as an input to the program written in C language and is run using the code composer studio. The results are obtained for different cases are shown below.

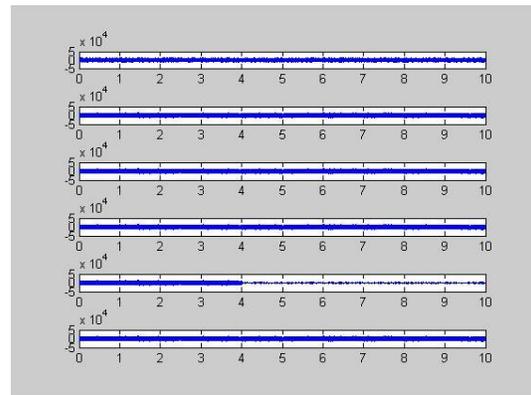

Fig.6. input data **1**

Comment: The six channel input data, indicating channel 5 is going bad (low) from duration 4 -10.

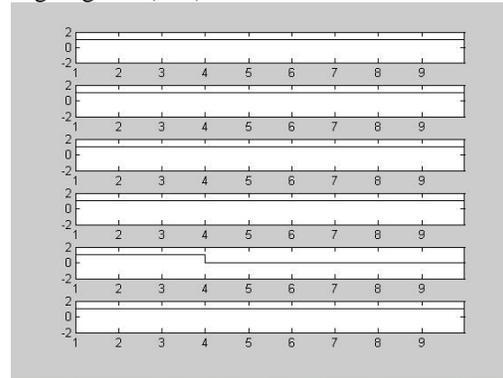

Fig.7. output data 1

Comment: The six channel output data, indicating all are good except channel 5 is going bad, which is reflected as low from the duration 4-10.

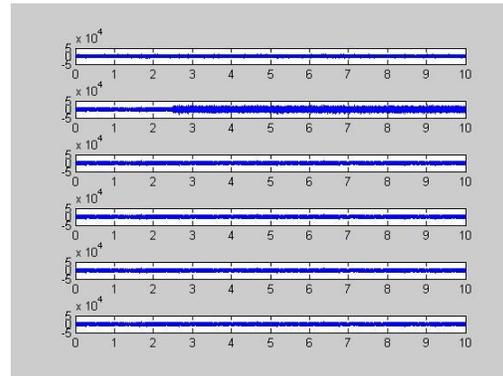

Fig.8. input data 2

Comment: The six channel input data, indicating channel 2 is going bad (high) from duration 2.5 -10.

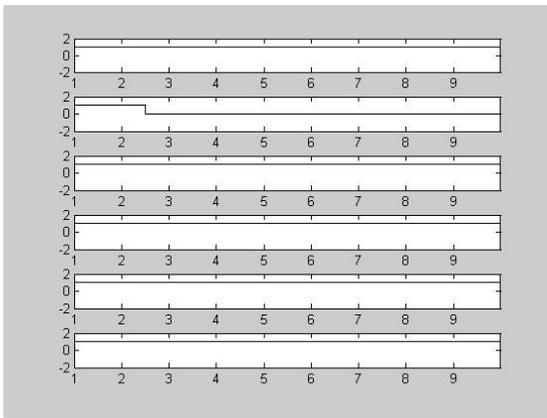

Fig.9. output data 2

Comment: The six channel output data, indicating all are good except channel 2 is going bad, which is reflected as low from the duration 2.5-10.

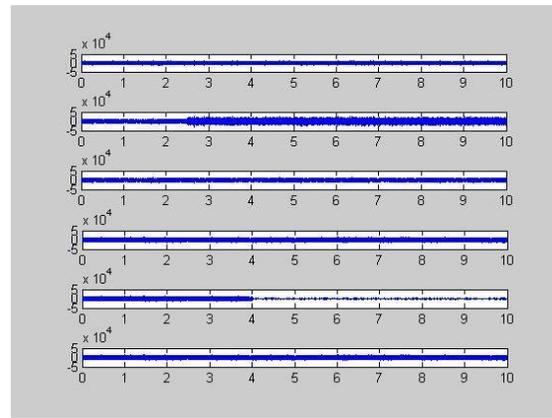

Fig.12. input data 4

Comment: The six channel input data, indicating channel 2 is going bad (high) from duration 2.5 -10, and channel 5 going bad (low) from duration 4-10.

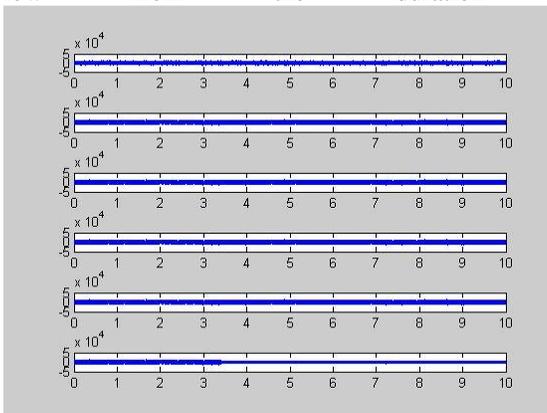

Fig.10. input data 3

Comment: The six channel input data, indicating channel 6 is going bad (low) from duration 3.5 -10.

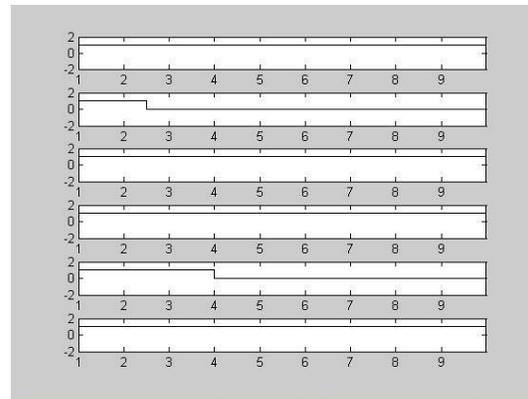

Fig.13. output data 4

Comment: The six channel output data, indicating all are good except channel 2 and channel 4 is going bad, which is reflected as low from the duration 2.5-10 for channel 2 and from the duration 4-10 for channel 4.

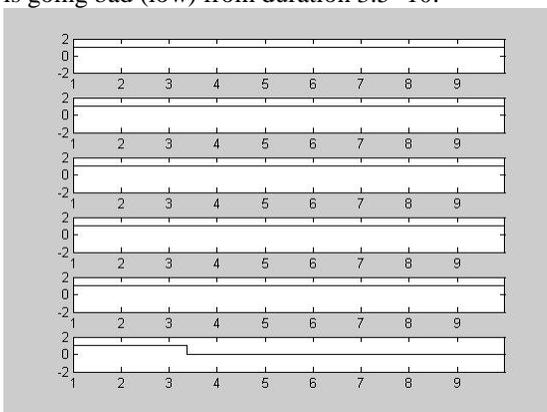

Fig.11. output data 3

Comment: The six channel output data, indicating all are good except channel 6 is going bad, which is reflected as low from the duration 3.5-10.

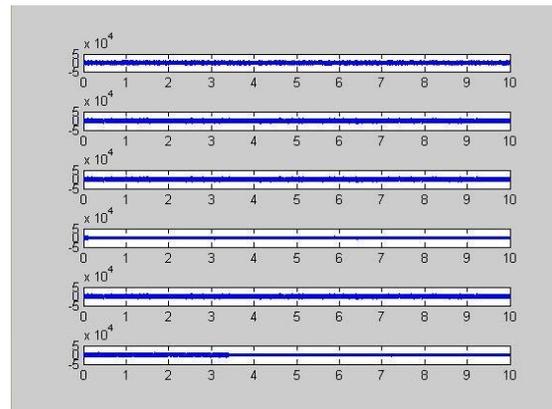

Fig.14. input data 5

Comment: The six channel input data, indicating channel 2 is going bad ( low ) from duration 1-10, and channel 6 going bad ( low) from duration 3.5-10.

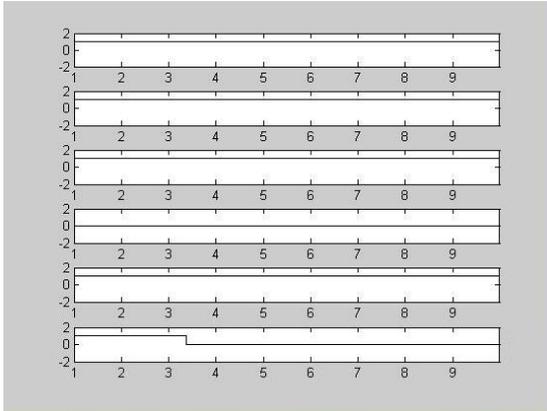

Fig.15. output data 5

Comment: The six channel output data, indicating all are good except channel 4 and channel 6 is going bad, which is reflected as low from the duration 1-10 for channel 4 and from the duration 3.5-10 for channel 6.

## IV. CONCLUSION

The data is compared with the results obtained. If we compare the output results with that of the data input, the output is becoming zero whenever there is less amplitude data in the input and also high amplitude in the input indicating the wrong data. The wrong data is identified and are displayed in the plots. Here if the two channel microphone data is bad then it will only be identified. The results are matching with the expected output. It proves that the algorithm implemented in C language is effectively working for the given data. It is successfully identifying and detecting the correct and wrong data. Hence we could verify and prove the redundancy software works better for achieving fault tolerance for testing of satellites with acoustic spectrum.

AUTHORS PROFILE

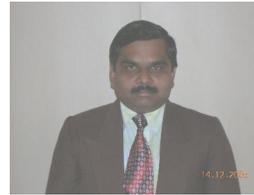

Sri. H.S. Aravinda Graduated in Electronics & Communication Engineering from University of Bangalore, India in 1997. He has a Post Graduation in Bio- Medical Instrumentation from University of My sore, India in 1999. His special interests in research are Fault tolerance, signal processing. He has been teaching engineering for UG & P G for last 12 years. He served various engineering colleges as a teacher and at present he is an Assistant professor in the Department of Electronics & Communication in Reva Institute of Technology & Management, Bangalore, India. He has more than 13 research papers in various National and International Journals & Conferences. He is a member of ISTE Also has served on the advisory and technical national conferences.

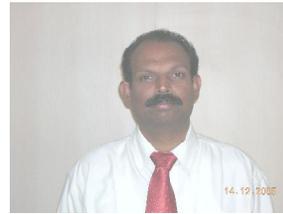

Sri. Dr. H.D. Maheshappa Graduated in Electronics & Communication Engineering from University of Mysore, India in 1983. He has a Post Graduation in Industrial Electronics from University of My sore, India in 1987. He holds a Doctoral Degree in Engineering from Indian Institute of Science, Bangalore, India, since 2001. He is specialized in Electrical contacts, Micro contacts, Signal integrity interconnects etc. His special interests in research are Bandwidth Utilization in Computer Networks. He has been teaching engineering for UG & P G for last 25 years. He served various engineering colleges as a teacher and at present he is a Professor & Head of the Department of Electronics & Communication in Reva Institute of Technology & Management, Bangalore , India. He has more than 35 research papers in various National and International Journals & Conferences . He is a member of IEEE, ISTE, CSI & ISOI. He is a member of Doctoral Committee of Coventry University UK. He has been a Reviewer of many Text Books for the publishers McGraw-Hill Education (India) Pvt., Ltd., Chaired Technical Sessions, and National Conferences and also has served on the advisory and technical national conferences.

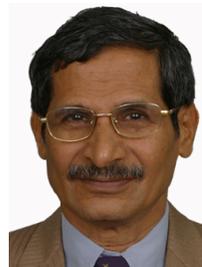

Sri. Dr.. Ranjan Moodithaya did his MSc from Mysore University in 1970 and got his Ph D from Indian Institute of Science in 1986. He joined NAL in 1973 after 2 yrs of teaching at Post Graduate centre of Mysore University. At present, he is Scientist G and he is incharge of the NAL-ISRO Acoustic Test Facility and NAL's Knowledge and Technology Management Division. He is a Life member of Acoustic Society of India and Aeronautical Society of India. His innovative products are sold to prestigious institutions like Westinghouse, Lockheed, Boeing and Mitsubishi through M/s. Wyle Laboratories, USA, the world leaders in the design of acoustic facilities. Dr. Ranjan has more than a dozen publications in international and national journals and more than 30 internal technical publications. He is a Life member of Acoustic Society of India and Aeronautical Society of India. He is also a Member of Instrument Society of India.